\documentclass[preprint,nofootinbib,eqsecnum,superscriptaddress,12pt]{revtex4-1}

\usepackage{color,graphicx,multirow,}

\usepackage{hyperref}
\hypersetup{
    colorlinks=true,
    linkcolor=blue,
    filecolor=magenta,      
    urlcolor=cyan,
}

\newcommand{\beq}{\begin{equation}}
\newcommand{\eeq}{\end{equation}}
\newcommand{\beqa}{\begin{eqnarray}}
\newcommand{\eeqa}{\end{eqnarray}}
\newcommand{\bea}{\begin{eqnarray}}
\newcommand{\eea}{\end{eqnarray}}
\newcommand{\bi}{\begin{itemize}}
\newcommand{\ei}{\end{itemize}}
\newcommand{\ben}{\begin{enumerate}}
\newcommand{\een}{\end{enumerate}}

\begin{document}

\title{CKM Substructure}
\author{Yuval Grossman}
\email{yg73@cornell.edu}
\affiliation{Department of Physics, LEPP, Cornell University, Ithaca,
NY 14853, USA \\[3mm]}
\author{Joshua T. Ruderman}
\email{ruderman@nyu.edu}
\affiliation{Center for Cosmology and Particle Physics, Department of
Physics, New York University, New York, NY 10003, USA\\[3mm]
\bigskip 
}

\begin{abstract}
The CKM matrix is not generic. The Wolfenstein parametrization encodes
structure by having one small parameter, $\lambda \approx 0.22$.  We pose the question: is there substructure in
the CKM matrix that goes beyond the single small parameter of the
Wolfenstein parameterization?
We find two relations that are approximately satisfied:
$|V_{td}|^2 = |V_{cb}|^3$
and
$|V_{ub}|^2  |V_{us}| = |V_{cb}|^4$.  We discuss the
statistical significance of these
relations and find that they may indicate deeper structure in the CKM
matrix.
The current precision, however, cannot exclude an $\mathcal{O}(10 \%)$ accident.
\end{abstract}

\maketitle

%%%%%%%%%%%%%%%%
\section{Introduction}

The CKM matrix is not generic. This fact is 
captured by the
Wolfenstein parameterization~\cite{Wolfenstein:1983yz}
\beq\label{eq:wolpar}
V=\pmatrix{
1-\frac12\lambda^2 & \lambda & A\lambda^3(\rho-i\eta)\cr
-\lambda  & 1-\frac12\lambda^2 & A\lambda^2 \cr
A\lambda^3(1-\rho-i\eta)&-A\lambda^2 & 1 \cr}   + \mathcal{O}(\lambda^4) \; .
\eeq
The assumption in writing the above approximation
is that $\lambda \approx 0.22$ is the only small
parameter, while $A$, $\eta$, and $\rho$ are all $\mathcal{O}(1)$.

In this paper we ask the following question: is there hidden structure in the
CKM matrix?   In other words, are there relations among CKM elements
beyond the hierarchies captured by the Wolfenstein parametrization?

A similar question has been asked regarding mixing in the lepton
sector where a clear hierarchy is not present and the PMNS matrix
is not close to the unit matrix. Is there structure in the PMNS matrix? While
we do not know the answer, two avenues have been explored. On the one
hand there are indications for structure, for example
tri-bimaximal mixing~\cite{Harrison:2002er}, and many models
have been proposed to explain it, notably
based on $A_4$ (for a review see, for example,
Ref.~\cite{King:2015aea}). An alternative framework is known as {\it anarchy}~\cite{Hall:1999sn,Haba:2000be,deGouvea:2003xe,deGouvea:2012ac,Lu:2014cla}. 
The PMNS matrix is defined to be anarchic by the authors of Ref.~\cite{Hall:1999sn}
 when ``all entries are comparable, no pattern or
structure is easily discernable, and there are no special precise
ratios between any entries.''

The question we would like to pose can be framed in a similar way to the question that was asked about the lepton sector.
To do this we define the term
{\it Wolfenstein anarchy}. A CKM matrix is Wolfenstein
anarchic if, besides the one small parameter $\lambda$, the matrix is otherwise
generic.  There should be no special precise relations among
parameters except those predicted by the Wolfenstein parameterization.
Is the CKM matrix found in Nature Wolfenstein anarchic?

The hierarchical nature of the CKM matrix motivates UV completions
that dynamically generate the small parameter $\lambda$.
For example, Froggatt--Nielson (FN)
models~\cite{Froggatt:1978nt,Leurer:1992wg,Leurer:1993gy,Ibanez:1994ig}
dynamically generate the parameter $\lambda$ as a ratio of mass
scales, and the special form of the CKM matrix is a consequence of
symmetries where fermions from different generations have different
charges.  These models predict hierarchies but only up to
$\mathcal{O}(1)$ numbers, which depend on $\mathcal{O}(1)$ parameters
within the UV completion.  Therefore, FN models predict that the CKM
matrix respects Wolfenstein anarchy.

In exploring whether the CKM matrix is consistent with Wolfenstein
anarchy we found two relations that hold to a good
approximation beyond the ones that are direct results of the
Wolfenstein hierarchy. They are
\beq \label{eq:rel-app}
|V_{td}|^2= |V_{cb}|^3, \qquad
|V_{ub}|^2= {|V_{cb}|^4 \over |V_{us}|} \, ,
\eeq
When combined, these relations lead to
\beq
\left|{V_{td} \over V_{ub}}\right|^2 = \left|{V_{us} \over V_{cb}}\right| \, .
\eeq
In term of the Wolfenstein parameters these relations are given by
\beq
A = \eta^2+(1-\rho)^2, \qquad
A^2 \lambda = \eta^2+\rho^2 \, .
\eeq
When combined we can write them as
\beq
{\eta^2+\rho^2 \over \eta^2+(1-\rho)^2} = A \lambda \, ,
\eeq
or as an $A$-independent relation
\beq \label{eq:A-ind-rel}
\left[\eta^2+(1-\rho)^2\right]^2 \lambda = \eta^2+\rho^2.
\eeq

There are two questions that we would like to ask about these
relations:
\begin{enumerate}
\item
What UV flavor models can generate them?
\item
Are they an indication of CKM substructure or are they 
consistent with Wolfenstein anarchy?
\end{enumerate}
Regarding the first question, we tried to find
a UV model that generates these relations, but were so far unable
to do so. The second question, on the other hand, is what we discuss below.

%%%%%%%%%%%%%%%%
\section{Fits of the CKM matrix elements}

We search for relations of the form 
\beq
 \prod_n |V_{{i_n}{j_n}}|^{a_n}  \prod_m |V_{{k_m}{l_m}}|^{-b_m} =1 \, , 
\eeq
where $i_n$ and $ k_m$ are selected from $(u,c,t)$, $j_n$ and  $l_m$ are selected from $(d,s,b)$, $V$ with any two subindices represents an element of the CKM matrix,
and $a_n$ and $b_n$ are positive integers.
We use the following criteria when looking for relations
\begin{enumerate}
\item
We check for relations that involve a small number of CKM elements,
which we choose to be six,
that is
\beq
\sum_n a_n + \sum_m b_m \le 6 \, .
\eeq
\item
We look for relations such that their central values are very close to
one, concretely, those that
fall within $2\%$ of one.
\item
We do not consider relations that are a result of the Wolfenstein 
parameterization, for example, $|V_{us}| = |V_{cd}|$ is not a new relation.
\end{enumerate}
Based on these criteria, and the central values of the CKM elements from
CKMfitter~\cite{Charles:2004jd}, we found exactly two independent relations, 
which are given by Eq.~(\ref{eq:rel-app}). 

\begin{figure}[t]
\begin{center}
\includegraphics[width=.65\textwidth]{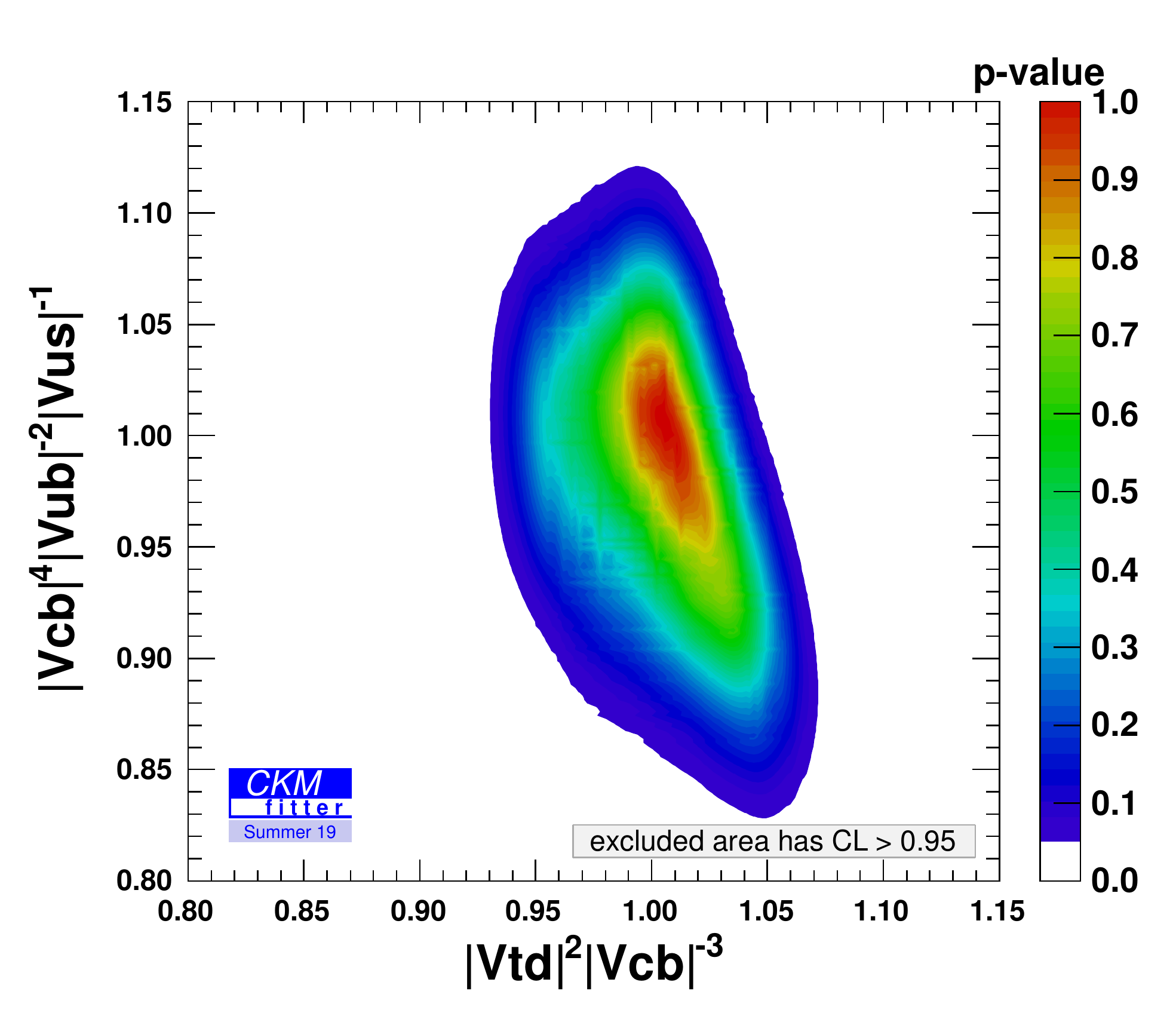}
\end{center}
\vspace{-.3cm}
\caption{\label{fig:cor-plot} \footnotesize
  The correlation between the two relations, $R_1$ and $R_2$, defined in Eq.~(\ref{eq:fit})~\cite{Sebastien}.}
\end{figure}

We
next assess the statistical significance of these
relations. The
CKMfitter code, modified to check for relations, is used to find~\cite{Sebastien}
\beq \label{eq:fit}
R_1 \equiv
\left|{V_{td}^2 \over V_{cb}^3}\right| =
1.004^{+0.026}_{-0.023} \, , \qquad
R_2 \equiv
\left|{V_{ub}^2  V_{us} \over V_{cb}^4}\right|  =
0.994^{+0.066}_{-0.044} \, .
\eeq
The input values and statistical procedure of this fit can be found in Ref.~\cite{Charles:2004jd}. 
Fig.~\ref{fig:cor-plot}. shows the same result as a correlated plot.

A few remarks are in order:
\begin{enumerate}
\item
The central values of both relations are within $1\%$ of 1.
\item
The errors in $R_2$ are larger basically since the error on $\left|V_{ub}\right|$
is the largest among the CKM elements (see for example the ``CKM Quark-Mixing Matrix'' review
in Ref.~\cite{PDG2020}). 
\item
The errors are asymmetric, which indicates that there are
significant correlations in deriving the errors and the errors are
not just simply statistical.
\item
There is no strong correlation in the statistical errors between the two relations.
\end{enumerate}

%%%%%%%%%%%%
%%%%%%%%%%%%
%%%%%%%%%%%%
%%%%%%%%%%%%

\section{A test for Wolfenstein Anarchy}

%%%%%%%%%%%%
%%%%%%%%%%%%
%%%%%%%%%%%%
%%%%%%%%%%%%
\begin{figure}[t]
\begin{center}
\includegraphics[width=1\textwidth]{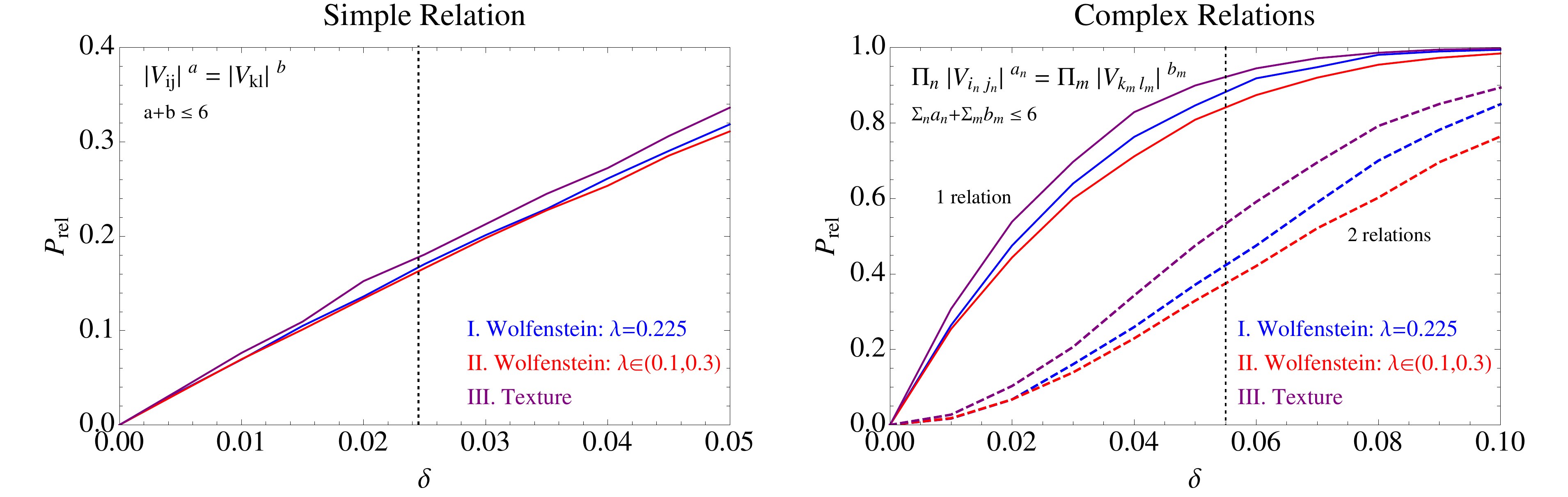}
\end{center}
\vspace{-.3cm}
\caption{\label{fig:prob}  \footnotesize
\emph{Left}: The probability of finding at least one simple relation,
as defined in Eq.~(\ref{eq:simple}), as a function of the precision,
$\delta$.  Random CKM matrices are generated using Wolfenstein Anarchy
with fixed $\lambda$ (blue), Wolfenstein anarchy with random $\lambda$
(red), and random Yukawa matrices using textures with random
coefficients (purple).  Parameters are randomly generated as described
in Eqs.~(\ref{eq:scanI}), (\ref{eq:scanII}), and (\ref{eq:scanIII}).     \emph{Right}:~The probability of finding
at least one (solid) or at least two (dashed) complex relations, as
defined in Eq.~(\ref{eq:complex}).  The colors correspond to different
scans, as on the left.  The vertical black lines at $\delta = 0.0245$
and $\delta=0.055$, on the left and right, respectively, correspond to
the (symmetrized) precision of the $R_{1,2}$ fits, see Eq.~(\ref{eq:fit}).
}

\end{figure}
%%%%%%%%%%%%
%%%%%%%%%%%%
%%%%%%%%%%%%
%%%%%%%%%%%%

%%%%%%%%%%%%
%%%%%%%%%%%%
%%%%%%%%%%%%
%%%%%%%%%%%%
\begin{figure}[t]
\begin{center}
\includegraphics[width=1\textwidth]{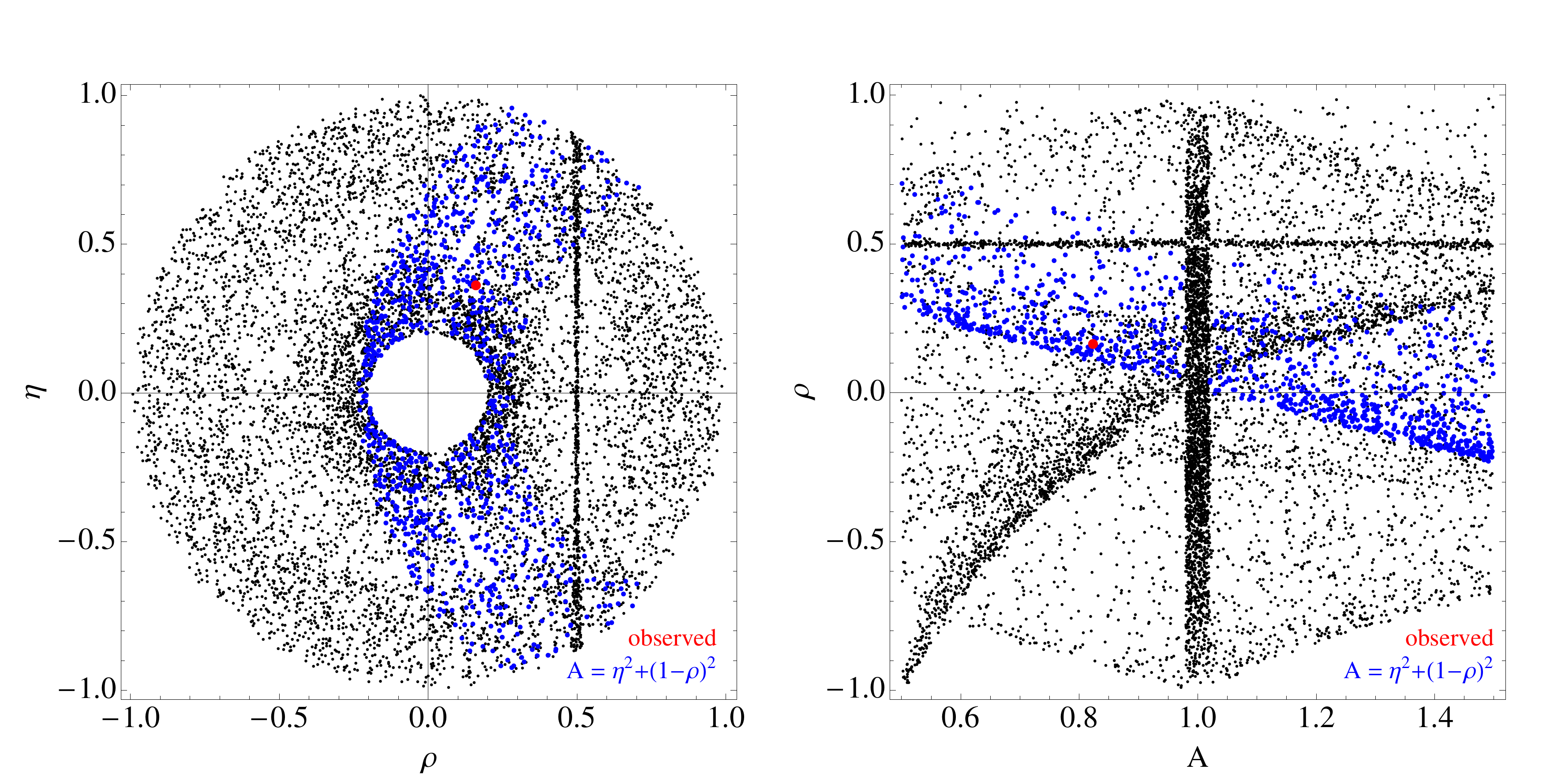}
\end{center}
\vspace{-.3cm}
\caption{\label{fig:dots} \footnotesize
The black dots show the $(\eta,\rho)$ (\emph{left}) and $(A,\rho)$
(\emph{right}) of $10^4$ CKM matrices generated with Wolfenstein
Anarchy that satisfy at least 1 simple relation, Eq.~(\ref{eq:simple})  with precision $\delta = 0.02$.  For this scan we fix $\lambda=0.225$.  We find that 13\% of random CKM matrices satisfy at least one relation.  The blue dots correspond to random CKM matrices that satisfy relation $R_1$, $A = \eta^2 + (1-\rho)^2$.  The red dot is the observed CKM matrix.  Several structures that are clearly visible in the plot correspond to relations that appear in Table~\ref{tab:relations}, such as $A=1$ and $\rho = 1/2$.}

\end{figure}
%%%%%%%%%%%%
%%%%%%%%%%%%
%%%%%%%%%%%%
%%%%%%%%%%%%

\begin{table}
\begin{center}
\begin{tabular}{c|c|c|c|c}
\multicolumn{2}{c|}{}  &  \multicolumn{3}{c}{~probability (\%)~}     \\ \hline
& & scan {\bf I} & scan {\bf II} & scan {\bf III} \\
~Wolfenstein~&~CKM~& ~$\lambda = 0.225$ ~& ~$\lambda  \in (0.1,0.3)$ ~& ~texture~ \\
\hline \hline
~$ A = 1$~ & ~$V_{cb} = V_{us}^2$~  & 3.7 &  3.7 & 0.8   \\
\hline
~$A \sqrt{(1-\rho)^2 + \eta^2} = 1$~& ~$V_{td} = V_{us}^3$~ & 2.6 & 2.6 & 0.7 \\ \hline
~$A \sqrt{\rho^2 + \eta^2} = 1 $~ &~$V_{ub} = V_{us}^3$~ & 1.9 & 1.9 & 0.6  \\ \hline
~$A =(1-\rho)^2 + \eta^2 $ ~&~$V_{td}^2 = V_{cb}^3$~ &  1.4 &  1.4 & 0.7 \\ \hline
~$A = \rho^2 +\eta^2 $ ~&~$V_{ub}^2 = V_{cb}^3$~& 1.0 & 1.0 &  0.6  \\ \hline
~$\rho = 1/2$~ &~$V_{ub} = V_{td}$~& 0.9 & 0.9 & 2.3  \\ \hline
~$A \sqrt{\rho^2 + \eta^2} = \lambda$~ &~$V_{ub}=V_{us}^4$~ & 0.8 & 0.7 & 0.6  \\ \hline
~$ A \lambda = \sqrt{\rho^2 + \eta^2}$~ &~$V_{ub}=V_{cb}^2$~ &  0.7 & 0.5 &  0.5 \\ \hline
~$A \sqrt{(1-\rho)^2 + \eta^2} = \lambda$~  &~$V_{td}=V_{us}^4$~ & 0.2 & 0.1 & 0.5 \\ \hline
~$A \lambda = \sqrt{(1-\rho)^2 + \eta^2}$~ & ~$V_{td}=V_{cb}^2$~& 0.1 & 0.1& 0.4  \\ \hline
\multicolumn{2}{c|}{~other~}    & 0.02 & 0.03 & 6.0 \\
\end{tabular}
\caption{\label{tab:relations} \footnotesize
We generated $10^6$ random CKM matrices using the prescriptions of
scans {\bf I}, {\bf II}, and {\bf III}, described by Eqs.~(\ref{eq:scanI}), (\ref{eq:scanII}), and (\ref{eq:scanIII}). We find that 13\%, 13\%, and 14\%, respectively, satisfy at least one simple relation, as in Eq.~(\ref{eq:simple}), with precision $\delta = 0.02$. We show the 10 most common relations for scan {\bf I}.  The first column shows each relation as a function of Wolfenstein parameters, and the second column is an example of how the relation manifests as a function of CKM matrix elements.   The final three columns show the probability that a random CKM matrix, from each scan, satisfies the relation with precision $\delta = 0.02$.  The last row shows the probability of satisfying additional relations not listed in this table.
}
\end{center}
\end{table}

Are the relations of Eq.~(\ref{eq:fit}) evidence of CKM substructure, or are they consistent with Wolfenstein anarchy?  This question can be addressed using a statistical model for Wolfenstein anarchy.  Given an ensemble of randomly generated CKM matrices that are consistent with the Wolfenstein parameterization, is it exceedingly unlikely to satisfy the relations of Eq.~(\ref{eq:fit})?  There is a ``trials factor" because there are many possible relations that could, in principle, be satisfied.  In order to assess the likelihood that the observed relations are random occurrences, we have used a computer to randomly generate many CKM matrices and checked the probability that they satisfy the above, or similar, relations.

We generate ensembles of random CKM matrices using two methods, described below.  First, we generate random Wolfenstein parameters directly.  Second, we generate random Yukawa matrices using textures, motivated by FN models, with random coefficients.  For alternate approaches to generating random CKM matrices see Refs.~\cite{Froggatt:1979sz, Rosenfeld:2001sc}.

\subsection{Random CKM Matrices from Wolfenstein Anarchy}

We generate ensembles of CKM matrices by choosing random Wolfenstein parameters according to two prescriptions, which we label as scan {\bf I} and scan {\bf II}.  For these scans we generate random Wolfenstein parameters with uniform (linear) probability distributions in the following ranges,
\beqa
({\rm \mathbf I}) & \quad \lambda = 0.225 \, , & \quad A \in (0.5-1.5)  \, , \quad \sqrt{\rho^2+\eta^2} \in (0.2-1) \, ,  \quad \theta \in (0,2\pi) \label{eq:scanI} \\ 
({\rm \mathbf{II}}) & \quad \lambda \in (0.1-0.3)  \, , & \quad A \in (0.5-1.5)  \, , \quad \sqrt{\rho^2+\eta^2} \in (0.2-1)  \, , \quad \theta \in (0,2\pi)  \label{eq:scanII}    \, \, ,
\eeqa
where we write $\rho + i \eta = \sqrt{\rho^2+\eta^2} \, e^{i \theta}$ and separately generate the magnitude and phase.  Scan {\bf I} fixes $\lambda$ to the observed value, while scan {\bf II} allows $\lambda$ to vary between 0.1 and 0.3.

We first look for CKM matrices satisfying a simple relation of the form:
\beq \label{eq:simple}
\left| |V_{ij}|^a |V_{jk}|^{-b} -1 \right| < \delta \qquad \qquad a+b \le 6 \, ,
\eeq
where $\delta$ is the precision by which the relation is satisfied.
This is a generalization of $R_1$ of Eq.~(\ref{eq:fit}). 
 The left of
Fig.~\ref{fig:prob} shows the probability that at least one simple
relation is satisfied as a function of the precision, $\delta$.  For
this test and the tests described below,
we generate $10^6$ random matrices, and we use all of them or a
subset, making sure that
that the statistical uncertainty is far smaller than the uncertainties in Eq.~(\ref{eq:fit}).  The
vertical dashed line corresponds to $\delta = 0.0245$, which is the
current precision fitting to $R_1$ after symmetrizing the
uncertainties, see Eq.~(\ref{eq:fit}).  
We find a probability of 16\% (15\%) that a random CKM matrix satisfies at least one relation to this precision for scan {\bf I} ({\bf II}).  

To further study the probability to have relations, we 
check the probability that a random matrix satisfies one or more relations with precision
$\delta = 0.02$. 
We find a probability of $13\%$ for this to be the case for both scans.  
Fig.~\ref{fig:dots} shows the locations of these
points in the $(\rho, \eta)$ and $(A, \rho)$ planes as black dots.
(To generate the plot we use a subset of $10^4$ of the matrices that
satisfy one or more relations.)
 We find that 1.4\% of random CKM matrices satisfy $R_1$, corresponding to the relation $A = \eta^2 + (1 + \rho)^2$.  These points are denoted by blue dots.  The red dot corresponds to the observed values of the Wolfenstein parameters.

While there is a percent-level chance that a random CKM matrix
satisfies $R_1$ to the observed precision, there is evidently an
order-of-magnitude larger probability that a similar relation is
satisfied.  What other relations are being satisfied, for example, by the black dots in Fig.~\ref{fig:dots}? Using the
generated random CKM matrices, the 10 most common relations are collected in Table~\ref{tab:relations}\@.  The table lists the probability that each relation is satisfied with a precision of  $\delta = 0.02$ for scans {\bf I} and {\bf II}. We observe similar probabilities for both scans.  Several of the relations are clearly visible as clusters of points in Fig.~\ref{fig:dots}, such as $A=1$ and $\rho=1/2$.  Relations not shown in the table together account for less than 0.02\% (0.03\%) of random matrices for scan {\bf I} ({\bf II}).

We next look for more complicated relations of the form,
\beq \label{eq:complex}
\left| \prod_n |V_{{i_n}{j_n}}|^{a_n} \prod_m |V_{{k_m}{l_m}}|^{-b_m} - 1 \right| <  \delta  \qquad \qquad   \sum_n a_n + \sum_m b_m \le 6 \, ,
\eeq
generalizing $R_2$ from above.  These relations include the simple relations of Eq.~(\ref{eq:simple}).  The right side of Fig.~\ref{fig:prob} shows the probability of satisfying 1 or 2 of these relations.  There is a large probability of satisfying one relation of this form because of the large number of possible relations.  For example the probability of satisfying one complex relation to a precision of $\delta = 0.02$ is 48\% (44\%) for $\lambda$ fixed (varied).  The probability that a randomly generated matrix satisfies two relations with $\delta = 0.02$ simultaneously is only 7\%.  The vertical dashed line shows $\delta = .055$, which is the current precision for fitting $R_2$.  We find that the probability of satisfying 2 relations with this precision is about 40\%.

\subsection{Random CKM Matrices from Textures}

The purpose of Wolfenstein anarchy is to capture the types of CKM matrices that we expect to be generated by UV completions that generate the hierarchical structure of the CKM matrix, such as Froggatt-Nielson~\cite{Froggatt:1978nt,Leurer:1992wg,Leurer:1993gy,Ibanez:1994ig}.  We would like to verify: does generating random Wolfenstein parameters produce similar results to generating random Yukawa matrices?

In order to answer this question we have also used a computer to generate random Yukawa matrices.  We assume the following textures~\cite{Leurer:1993gy},
\beq \label{eq:texture}
Y_d = 
\left(\begin{array}{ccc}
c_{11} \lambda^6  & c_{12} \lambda^5 & c_{13} \lambda^5 \\ 
c_{21}\lambda^5 & c_{22} \lambda^4  & c_{23} \lambda^4 \\ 
c_{31}\lambda^3 & c_{32} \lambda^2 &  c_{33} \lambda^2 \end{array}\right)
\qquad Y_u = 
\left(\begin{array}{ccc}
d_{11} \lambda^6 & d_{12}  \lambda^4 &  d_{13} \lambda^3  \\
d_{21} \lambda^5 & d_{22} \lambda^3 &  d_{23} \lambda^2 \\
d_{31} \lambda^3 & d_{32} \lambda  & d_{33}  \end{array}\right)
\eeq
where $\lambda = 0.225$ and $c_{ij}$ and $d_{ij}$ are random $\mathcal{O}(1)$ constants.  For alternate textures that could also be considered, see for example Ref.~\cite{Ibanez:1994ig}.  

Scan {\bf III} is defined by selecting the following parameters with uniform (linear) probability distributions,
\beqa
({\rm \mathbf {III}}) & \qquad \lambda = 0.225  & \qquad  |c_{ij}|, |d_{ij}| \in (0.2-5) \qquad   \arg(c_{ij}), \arg(d_{ij})\in (0,2\pi) \label{eq:scanIII}  \, \, .
\eeqa
Note that a specific UV completion may predict correlations among the various coefficients, but for simplicity we do not include any correlations here.

Fig.~\ref{fig:prob} shows the probability of satisfying simple and complex relations from random textures, compared to the  Wolfenstein anarchy approach described above.  We find similar probabilities in both cases, confirming that the simpler Wolfenstein anarchy approach captures the behavior of random Yukawa matrices built from hierarchical textures.  

We also generate $10^6$ random Yukawa matrices using the prescription of scan {\bf III}, and look for simple relations of the form of Eq.~(\ref{eq:simple}) with a precision
$\delta = 0.02$.  We find that 14\% of random matrices  in scan {\bf III} satisfy one or more relation with this precision.  This is similar to the 13\% rate, found above, for scans {\bf I} and {\bf II}.  Table~\ref{tab:relations} shows the probability of satisfying various relations for all three scans.  Interestingly, although the overall rate of satisfying one or more relations is similar across all three scans, scan {\bf III} exhibits a different pattern of relative probabilities for satisfying the various relations.

\section{Discussion and conclusions}

The main result of our paper is Eq.~(\ref{eq:fit}) where we
point out two relations between CKM parameters.  We are
interested in them because they may point towards a deeper structure that
generates the flavor parameters of the SM\@.
Using the current values of the CKM parameters,
the relations hold to a precision on the order of a few percent.
This observation led us to define the notion of
``Wolfenstein anarchy'' where the assumption is that there is one
small parameter, $\lambda$, and there is no additional structure in
the quark flavor sector.  We then perform a simple statistical
analysis to find out what is the probability that such relations are a
result of Wolfenstein anarchy.  We find that with current uncertainties on the
values of the parameters, and accounting for the trials factor of satisfying other similar relations, the probability is not $\mathcal{O}(1)$, but is also not small,
that is, the probability is above $10\%$.

In order to make progress testing our relations, we require
advances in both the theoretical and experimental inputs for the extraction of
CKM parameters. In a few years, we expect that CKM fits will achieve the precision to either show that our relations are violated, or to confirm that they hold with a precision to disfavor the ansatz of Wolfenstein anarchy.
 At present, we think that it is too early to make this call,
and therefore our purpose in this paper is simply to point out these relations and to suggest a simple framework for assessing their significance.

Note that the CKM elements
run~\cite{Balzereit:1998id,JuarezWysozka:2002kx}. To a good
approximation only $A$ runs. The above two papers do not agree on
the exact magnitude of the effect. In the SM, they found that the value of $A$ is
reduced by
$13\%$~\cite{Balzereit:1998id} ($25\%$~\cite{JuarezWysozka:2002kx})
from the weak scale to the GUT scale.  The running can be
different in cases that there are new fields, beyond the SM, with masses below the scale where the flavor
physics is generated. If the relations we studied
here do come from some fundamental UV physics, what is important are
the values of the CKM elements at the scale where this UV physics is
generated.  Assuming the SM  and given the current $2\sigma$ errors on the
values of the parameters, this implies that the
relations $R_1$ and $R_2$ can be satisfied at a scale that
is roughly below $10^6$ ($10^5$) GeV using
Ref.~\cite{Balzereit:1998id}
(Ref.~\cite{JuarezWysozka:2002kx}). The relation we found in
Eq.~(\ref{eq:A-ind-rel}) is independent of $A$ and thus is to a very
good approximation independent of the scale where the flavor structure
is generated.

Assuming that the relations we found do come from a UV theory, this
imposes a model-building challenge. In fact, we were unable to construct
a model to explain either of the relations. The main challenge is that
most UV models of flavor have more parameters in the UV than in the
IR\@.  In our case, however, we encounter the opposite situation: our IR relations suggest that the UV theory should have fewer parameters.

Finally, we remark that there could in principle be more relations beyond the types that we search for.  In particular, we only look for relations of the
form $a-b=0$, where $a$ and $b$ are the products of a few magnitudes of CKM
elements.
One can envision relations with more than two terms, with other proportionality
factors between terms, or with dependence on the phase.

While the current precision does not allow us to claim that the relations of Eq.~(\ref{eq:fit}) hint at a deep fact about UV
physics, they motivate searching for UV models that can explain them.

\section*{Acknowledgments}
We thank S\'ebastien Descotes-Genon for useful discussions and for
performing the fit of Eq.~(\ref{eq:fit}) and
Fig.~\ref{fig:cor-plot}.
We thank the people of Kibutz Neot Semadar in Israel for
hospitality and for inspiring us to look for unexpected relationships.
The work of YG is supported in part by the NSF grant PHY1316222\@. JTR
is supported by NSF CAREER grant PHY-1554858 and NSF grant PHY-1915409.

%%%%%%%%%%%%

\end{document}